# AGES OF GALAXY BULGES AND DISKS FROM OPTICAL AND NEAR-INFRARED COLOURS

R. F. PELETIER
*Kapteyn Astronomical Institute*
*Postbus 800, 9700 AV  Groningen, Netherlands*
*and*
*Instituto de Astrofísica de Canarias*
*Via Lactea s/n, 38200 La Laguna, Tenerife, Spain*

AND

M. BALCELLS
*Kapteyn Astronomical Institute*
*Postbus 800, 9700 AV  Groningen, Netherlands*

**Abstract.** For a sample of bright nearby early-type galaxies we have obtained surface photometry in bands ranging from U to K. Since the galaxies have inclinations larger than $50^o$ it is easy to separate bulges and disks. By measuring the colours in special regions, we minimize the effects of extinction, and by looking at $B - K$ colour gradients we can show that for these type of spirals the colours mainly give information about stellar populations, and not extinction. We find that the differences between bulges and disks in all colours is very small, and using simple population models we can show that on the average the age difference between the bulge and the disk at 2 scale length is smaller than 30%, and much smaller if part of the difference is caused by a gradient in metallicity.

## 1. Introduction

Age differences between bulges and disks can provide important basic checks on theories of the assembly of disk galaxies. What do galaxy colours tell us about the age of formation of bulges and disks? According to the classical concept of Population, disks should be Population I and young, while bulges belong to Population II and are old. Galaxy photographs give the same picture. To address this question, external galaxies might give better clues than our Galaxy, due to the large obscuration towards the Bulge. The drawback is still that one has to analyze integrated colours or spectra.

Here we discuss optical and near-infrared colours. As has been shown by e.g. Frogel (1985) age and metallicity can be disentangled by using a blue optical colour index as a measure of recent star formation, and a near-infrared colour index as a measure of metallicity. Bothun & Gregg (1990) applied this method to S0 galaxies. They found that for a given $J - K$ disks are generally bluer by $\sim$0.4 mag in $B - H$ than bulges, corresponding to an age difference of more than 5 Gyr. Their infrared measurements however are from off-nuclear aperture photometry, which is extremely complicated, and can produce large errors. Here we use near-infrared images, combined with optical CCD-data, allowing us to obtain high-quality measurements. We present some colour-colour relations, and concentrate on the determination of ages. More details are given in Balcells & Peletier (1994, **BP**) and Peletier & Balcells (1995).

## 2. The sample, observations and the determination of colours

The sample discussed here is a diameter-limited sample of inclined galaxies (i $\geq$ 50º) of type S0 - Sbc. Except for 2 objects, the sample is the same as described by BP, where the optical data for the bulges are presented. The sample was observed at $K$ at UKIRT with IRCAM3, equipped with a 256 $\times$ 256 InSb array. 20 of the 30 galaxies were also observed at $J$. Mosaics were made with a total size of 100" $\times$ 100", a pixel size of 0.291" and an effective seeing between 0.8" and 1".

Colour profiles were obtained in wedges centered on the center of the galaxy in $K$. For the bulges these wedges were centered on the minor axis, with a width of 45º. For the disks, a 10º wide wedge was taken 15º away from the major axis. In both cases, colours were obtained on the non-dusty side, this way minimizing the effects of extinction. Given the orientation of these galaxies, it happens that the bulge profiles, except sometimes very close to the center, are really dustfree, and only suffer from disk light behind the bulge. Given that the surface brightness of bulges is larger than that of disks, disk light behind the bulge does not affect the colours. Some disk-profiles, even on the non-dusty side, are still rather dusty. However, scale length ratios between $B$ and $K$ in most of our galaxies are close to 1, and only the ones in some Sb's and Sbc's are similar to the ratios in the galaxies of Peletier *et al.* (1995), so that we can still assume that, in most cases, the effects of extinction are unimportant.

## 3. Colours of bulges and disks

Galaxy disks and bulges have negative colour gradients, i.e. colours become bluer radially outward (de Jong 1995; BP) , but the gradients are small enough that we could assign representative values for the colour of each component. For bulges, we have taken the colour at $0.5 \times r_{eff}$ or at 5 arcsec, whichever is larger. For disks we use the colours at 2 major axis scale lengths. We find that disk colours, while somewhat bluer, are very similar to bulge colours for all the galaxies (see Fig. 1). Here the diagonal line indicates the locus where both colours are equal. The average differences between disk and bulge and bulge colours are $0.126 \pm 0.165$ for $U - R$, $0.045 \pm 0.097$ for $B - R$, $0.078 \pm 0.165$ for $R - K$ and $0.016 \pm 0.087$ for $J - K$. Unless there is a conspiracy between metallicity and age, this diagram implies that age differences between bulges and disks are small. This information however is only useful if we also know the ages themselves. We can obtain this information from Fig. 2, where we plot $R - K$ of the bulges vs. $U - R$, analogous to Bothun & Gregg (1990). Also plotted are single age - single metallicity models by Vazdekis *et al.* (1995) for 17, 12, 8, 4 and 1 Gyr. We estimate the systematic error in the model in the colours to be $\sim 0.2$ mag, by comparing models from various sources in the literature. Especially the U-band and the near-infrared are hard to synthesise, the first because of lack of good calibration stars, and the second because of uncertainties in advanced stages of stellar evolution. It would be *definitely wrong* to use this diagram to age-date these bulges. What can be done however is say that there is a large range in ages for the bulges. Note also the scale. The range in colours between the galaxies is much larger than the average colour difference between bulge and disk of the same galaxy.

Using the same models, we now quantify the colour differences between bulges and disks in terms of age differences (in case of constant metallicity), or metallicity differences (in case of constant age). Since it is likely, from e.g. abundance measurements in HII regions, that there are metallicity gradients, the real situation will probably be somewhere in between. For the sample, we find that the average difference in log Z will be $\sim 0.12 \pm 0.16$ $(1\sigma)$, and $\sim 0.11 \pm 0.15$ in $\log(t)$. The numbers imply that a bulge is at most 30% older than its disk, or much less, if metallicity gradients are taken into account. Some bulges are even slightly bluer than the corresponding disk (but the same colour within the errors), and in these cases one would infer the same age for both.

## 4. Discussion

The data of this paper put the strong constraint that disks of early-type spirals must have been together with, or just after the bulge. It does not endanger the traditional model of Eggen, Lynden-Bell & Sandage (1963) of galaxy formation, but on the other hand implies that it is very unlikely that there were three discrete events of massive star formation: for the bulge, the disk, and the thick disk. The properties of bulge and inner disk seem to merge smoothly in terms of for example kinematical properties and kind of stars. The continuous infall model of Gunn (1982), where the age of the stars is determined by the free-fall collapse time of the infalling gas, would produce such a system, since these time scales will be similar for bulge and inner disk.

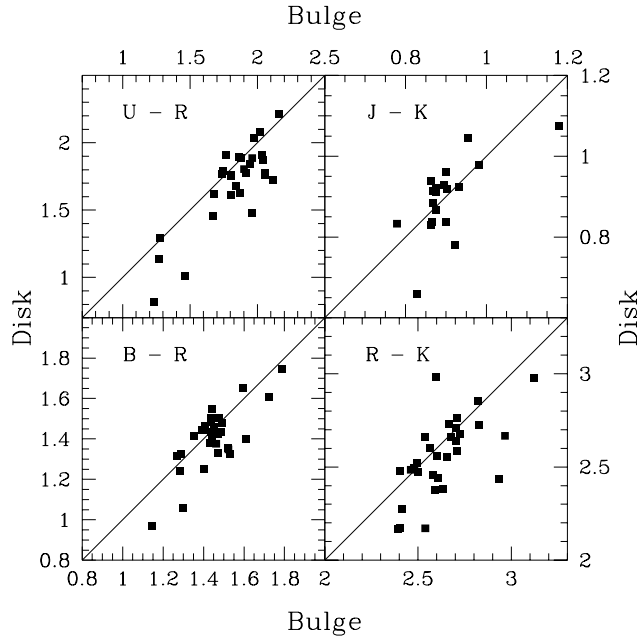

Figure 1. Disk colours as a function of bulge colours

An alternative way of forming bulges has been proposed by Combes *et al.* (1990) and others. Bulges may appear through the formation and later desctruction of a bar due to instabillities in a dynamically cold disk. This method does not predict any age differences, and is in good agreement with our results here. However, galaxies with large bulges like the Sombrero cannot be formed in this way, and also one would not expect a continuous change in colours, colour gradient and surface brightness profile shape from large bulges of S0's to small Sb bulges (Andredakis *et al.* 1995).

To summarise, we have found that colour differences in many optical and near-infrared colours between bulges and disks at 2 scale lengths are very small, implying very small age differences. Our result is in disagreement with Bothun & Gregg's (1990) result for S0 galaxies.

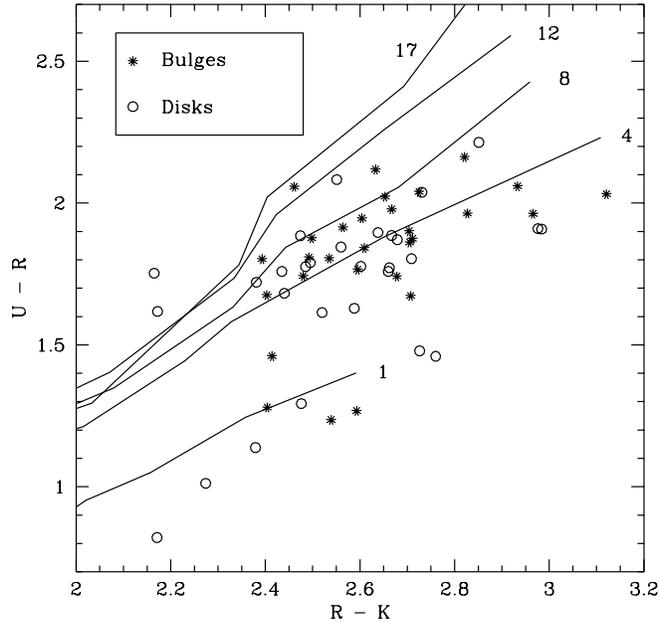

*Figure 2.*  *U − R* vs. *R − K* relations for bulge colours. Also included are single age, single metallicity models of Vazdekis *et al.* (1995)